\documentclass{article}
\usepackage{amsmath, amssymb, amsthm,bbm}
\usepackage{algorithm, algorithmic}
\usepackage{geometry}

\RequirePackage[numbers]{natbib}

\usepackage{orcidlink}

\theoremstyle{plain}
\newtheorem{theorem}{Theorem}
\newtheorem{assume}{Assumption}
\newtheorem{lemma}{Lemma}

\title{Heavy Lasso: sparse penalized regression under heavy-tailed noise via data-augmented soft-thresholding}
\author{The Tien Mai\orcidlink{0000-0002-3514-9636}}

\date{
\small
Norwegian Institute of Public Health, Oslo, 0456, Norway
\\
email: the.tien.mai@fhi.no
}

\begin{document}

\maketitle

\begin{abstract}
High-dimensional linear regression is a fundamental tool in modern statistics, particularly when the number of predictors exceeds the sample size. The classical Lasso, which relies on the squared loss, performs well under Gaussian noise assumptions but often deteriorates in the presence of heavy-tailed errors or outliers commonly encountered in real data applications such as genomics, finance, and signal processing. To address these challenges, we propose a novel robust regression method, termed Heavy Lasso, which incorporates a loss function inspired by the Student’s t-distribution within a Lasso penalization framework. This loss retains the desirable quadratic behavior for small residuals while adaptively downweighting large deviations, thus enhancing robustness to heavy-tailed noise and outliers. Heavy Lasso enjoys computationally efficient by leveraging a data augmentation scheme and a soft-thresholding algorithm, which integrate seamlessly with classical Lasso solvers.
Theoretically, we establish non-asymptotic  bounds under both 
$\ell_1$
  and 
$\ell_2 $	
norms, by employing the framework of localized convexity, showing that the Heavy Lasso estimator achieves rates comparable to those of the Huber loss.
Extensive numerical studies demonstrate Heavy Lasso’s superior performance over classical Lasso and other robust variants, highlighting its effectiveness in challenging noisy settings.
\end{abstract}

Keywords: robust regression; Lasso; heavy-tailed noise; soft-thresholding; non-asymptotic bounds

\section{Introduction}
\label{sc_introduc}

High-dimensional linear regression has become a cornerstone of modern statistical methodology \citep{hastie2009elements,buhlmann2011statistics,giraud2021introduction}, particularly in applications where the number of covariates $p$ is large—often exceeding the sample size $n$. 
In such settings, the classical least squares estimator becomes ill-posed, motivating the use of regularization techniques to enforce sparsity and stabilize the estimation procedure. 
Among these, the least absolute shrinkage and selection operator (Lasso) introduced by \cite{tibshirani1996regression} has emerged as one of the most popular and computationally tractable approaches for high-dimensional regression and the package ``\texttt{glmnet}" has become popular for implementing Lasso \citep{glmnetpackage}. The Lasso estimator solves the optimization problem:
$$
\widehat{\boldsymbol{\beta}}^{\text{lasso}} = \arg\min_{\boldsymbol{\beta} \in \mathbb{R}^p} \left\{ \frac{1}{2n} \| \mathbf{y} - \mathbf{X} \boldsymbol{\beta} \|_2^2 + \lambda \| \boldsymbol{\beta} \|_1 \right\},
$$
where $\mathbf{y} \in \mathbb{R}^n$ is the response vector, $\mathbf{X} \in \mathbb{R}^{n \times p}$ is the design matrix, and $\lambda > 0$ is a tuning parameter that controls the amount of shrinkage. The success of the Lasso in variable selection and prediction has led to an extensive theoretical and empirical literature analyzing its statistical properties under various assumptions on the design matrix and error distribution, see for example \cite{zou2006adaptive,bunea2007sparsity,zhao2006model,lounici2008sup,bellec2018slope}.

A central motivation for the classical Lasso  is that the noise vector $\boldsymbol{\varepsilon} \in \mathbb{R}^n$, defined through the linear model 
$$
\mathbf{y} = \mathbf{X} \boldsymbol{\beta}^* + \boldsymbol{\varepsilon}
,
$$
has Gaussian-type noise which leads to the squared loss. However, in many real data applications---including genomics, finance, signal processing, and environmental statistics---the assumption of light-tailed noise is violated. 
Data may be contaminated with outliers, or the underlying noise distribution may be inherently heavy-tailed, such as a Student’s $t$-distribution or a mixture model with impulsive components. 
In such cases, standard Lasso estimators can suffer from substantial degradation in performance, both in terms of parameter estimation and variable selection \citep{loh2017statistical,loh2024theoretical}.

These challenges have spurred the development of Lasso-type methods designed to withstand the effects of outliers and heavy-tailed noise distributions. 
A growing body of work tackles this issue by replacing the traditional squared error loss with more robust alternatives. Notably, many recent approaches adopt
the Huber loss \citep{rosset2007piecewise,dalalyan2019outlier,loh2017statistical,loh2021scale} 
or Tukey’s biweight loss \citep{chang2018robust,smucler2017robust} to mitigate the influence of extreme observations. Other strategies include regression using
the $\ell_1 $ loss \citep{wang2007robust}, 
density power divergence methods \citep{zhang2010nearly,ghosh2020ultrahigh}, 
and generalized $\ell_q$ losses with $1 \leq q < 2$ \citep{wang2022high}. In addition, 
rank-based methods \citep{rejchel2020rank,wang2020tuning,zhou2024sparse} and median-of-means techniques \citep{lecue2020robust} have emerged as powerful tools for enhancing robustness in high-dimensional estimation. Other robust approach includes \cite{bhatia2015robust}. For a recent comprehensive overview of theoretical advances in this area, see \cite{loh2024theoretical}.

In this work, we introduce a novel robust loss function tailored for high-dimensional linear regression in the presence of heavy-tailed noise and potential outliers, within a Lasso penalization framework. We call our proposed method as ``Heavy Lasso". The core innovation lies in replacing the conventional squared loss with a robust surrogate based on the Student-t distribution. This surrogate retains the favorable properties of the squared loss for small residuals while adaptively downweighting the influence of large deviations, thereby enhancing robustness. A key advantage of the proposed method is its computational feasibility: the estimator remains efficiently implementable via a single additional data augmentation step compared to the classical Lasso. Specifically, we develop a data-augmentation-based soft-thresholding algorithm that seamlessly integrates with standard Lasso procedures to compute the robust estimator efficiently.

The proposed Heavy Lasso procedure not only offers empirical robustness but is also supported by strong theoretical guarantees. Our analysis builds on the powerful framework of localized convexity---a modern technique introduced by \cite{loh2017statistical}---which enables theoretical control over a broad class of loss functions that may be non-convex globally but exhibit favorable curvature properties in restricted local neighborhoods around the truth. By leveraging this framework, we derive non-asymptotic upper bounds on the estimation error of our method under both the $\ell_1$ and $\ell_2$ norms, even in high-dimensional regimes where the number of predictors exceeds the sample size. These bounds mirror those established for robust M-estimators such as the Huber loss in previous works \cite{loh2017statistical,loh2021scale}, affirming that our Student-inspired loss enjoys similar theoretical benefits. Importantly, these results hold under minimal moment conditions on the noise distribution, making our estimator particularly well-suited for settings involving heavy tails and outliers.

To demonstrate the effectiveness of our method, we conduct extensive simulation studies under a variety of settings involving heavy-tailed noise and outliers. We compare our approach to several Lasso variants that employ different loss functions, including the squared loss \citep{tibshirani1996regression}, $\ell_1$ loss \citep{wang2007robust}, Huber loss \citep{yi2017semismooth}, and a rank-based loss \citep{wang2020tuning}. The simulation results indicate that our method consistently exhibits strong empirical performance relative to these alternatives, particularly in challenging settings with non-Gaussian errors.
In addition to simulations, we present a real data application that further supports the utility of our approach. In this example, our method performs comparably to the Huber loss, illustrating its robustness and practical relevance in real data applications.

The remainder of the paper is organized as follows. Section \ref{sc_model_method} introduces the regression model along with our novel robust loss function, and provides theoretical guarantees for the proposed method. In Section \ref{sc_algorithm}, we describe our innovative soft-thresholding iterative algorithm designed to efficiently compute the estimator. Section \ref{sc_simulations} presents extensive simulation studies comparing our approach with four state-of-the-art methods. Section \ref{sc_real_data} focuses on a real data application to demonstrate practical utility. Finally, Section \ref{sc_conclusion} offers a discussion and concluding remarks. Technical proofs are deferred to the Appendix \ref{sc_proof}. Our method is implemented in the \texttt{R} package \texttt{heavylasso} available on Github: \url{https://github.com/tienmt/heavylasso} .

\section{Problem and Method}
\label{sc_model_method}

\subsection{Model and Method}
Throughout the paper, we assume that the observations $\{(x_i, y_i)\}_{i=1}^n$ are independently and identically distributed (i.i.d.) samples generated from the linear regression model:
\begin{equation}
\label{eq_linear_model}
  y_i = x_i^\top \beta^* + \epsilon_i,   
\end{equation}
where \(x_i \in \mathbb{R}^p\) is the \(i\)-th row of the design matrix \(X\); 
\(\beta^* \in \mathbb{R}^p\) is the coefficient vector. 
Here, we consider the 0-mean noise \( \epsilon_i\). 

We assume that the support size $s := |\operatorname{supp}(\beta^*)|$ is small, $ s <n <p $.

The standard Lasso estimator assumes Gaussian noise and minimizes squared error with an \(\ell_1\) penalty. However, in the presence of heavy-tailed noise, squared loss leads to inefficient and biased estimates. Instead, we propose minimizing the following loss function penalized by the \(\ell_1\)-norm:
\begin{equation}
\widehat{\beta} = \arg \min_{\beta \in \mathbb{R}^p} \left\{
\frac{ 1}{2n} \sum_{i=1}^n \ell_i (\beta) 
+
\lambda \|\beta\|_1
\right\},
\label{eq:student_t_lasso}
\end{equation}
\begin{equation}
\label{eq_lossfunction}
\text{where } \quad
\ell_i (\beta)
    :=
(\nu + 1 ) \log \left( 1 + \frac{(y_i - x_i^\top \beta)^2}{\nu } \right)   
,
\end{equation}
here, $ \lambda >0 $ is a tuning parameter.

The estimator defined in \eqref{eq:student_t_lasso} is called the \textbf{Heavy Lasso} estimator, and we refer to the robust loss function in \eqref{eq_lossfunction} as the \textit{Student loss function}.

\subsection{Intuition on the novel robust loss}
The loss function in \eqref{eq_lossfunction} behaves like a robustified version of the squared loss, similar in spirit to the Huber loss, because:
\begin{itemize}
    \item For small residuals $|y_i - x_i^\top \beta|$, we have that $\ell_i(\beta) \approx (\nu +1)\frac{(y_i - x_i^\top \beta)^2}{\nu }$, quadratic near zero.
    \item For large residuals, the growth is logarithmic, so less sensitive to outliers.
    
\end{itemize}
The loss function presented in \eqref{eq_lossfunction} is inspired from the negative log-likelihood under the assumption that the observational noise follows a Student’s $t$-distribution with $\nu > 0$ degrees of freedom. However, we remind that we do not assume specifically that the noise in our model \eqref{eq_linear_model} is a Student noise. This formulation naturally leads to a robust objective that accommodates heavy-tailed residuals. For $\nu =1$, one recovers the Cauchy loss function \citep{loh2017statistical,loh2024theoretical}.
Furthermore, the flexibility of the loss can be enhanced by introducing a positive scaling constant $c > 0$, resulting in the generalized form
$
\ell_i(\beta) = c(\nu + 1) \log \left(1 + \frac{(y_i - x_i^\top \beta)^2}{\nu} \right).
$
Such a modification preserves the essential structure of the loss and does not alter the theoretical properties or the convergence behaviour of the proposed algorithm. This scaled variant may offer practical advantages, such as tuning robustness relative to the scale of the data, while retaining the same robust penalization of large residuals inherent in the original formulation.

\begin{figure}[!ht]
    \centering
    \includegraphics[width=12cm]{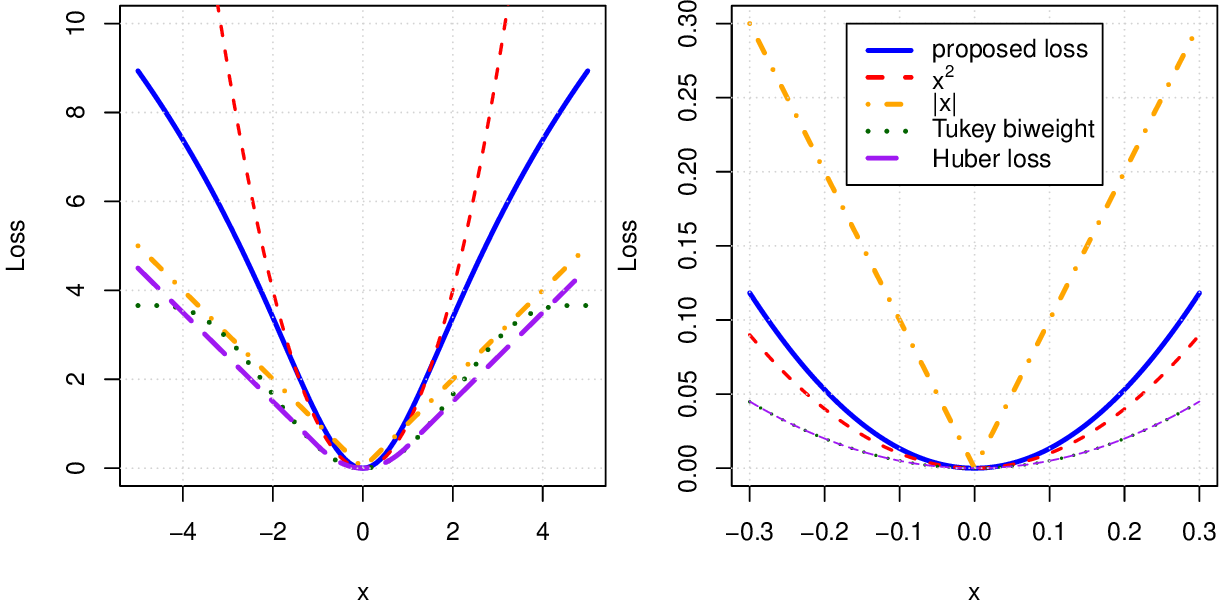}
    \caption{Comparison of our proposed loss function with other common losses: squared loss, absolute ($\ell_1$) loss, Tukey’s biweight loss, and Huber loss. The plot illustrates that, unlike the squared loss, our loss is much less sensitive to large residuals while closely resembling the squared loss for small residual values. Right: zoomed-in view near zero residuals. Left: full-scale plot.
    }
    \label{fig_compare_robustloss}
\end{figure}

Figure \ref{fig_compare_robustloss} presents a comparative analysis of the proposed loss function against several widely used alternatives, namely the squared loss, absolute loss, Huber loss, and Tukey’s biweight loss. The figure illustrates that, in contrast to the squared loss—which penalizes large residuals quadratically and is therefore highly sensitive to outliers—our loss function demonstrates a significantly more tempered response to large residuals. This robust behaviour ensures that extreme values exert less influence on the estimation process, enhancing the model’s stability in the presence of heavy-tailed or contaminated data. At the same time, for small residual values, our loss function exhibits a quadratic form similar to that of the squared loss, thereby preserving efficiency in the region where the Gaussian assumption is approximately valid. This dual behaviour closely mirrors the adaptive nature of the Huber loss, which also transitions from a quadratic regime for small residuals to a linear regime for larger ones. Overall, the plot underscores that our loss function retains the desirable robustness properties of Huber and Tukey losses while maintaining computational tractability and theoretical coherence in the context of robust regression.

\subsection{Theoretical guarantee}
We now demonstrate that, under suitable assumptions, our Heavy Lasso method achieves strong non-asymptotic theoretical guarantees comparable to those established for the Huber loss.

\begin{assume}
\label{assum_main}
Assume:
\\
 (A1): $\|x_i\|_\infty \le K$ for all $i$,
\\
 (A2): the empirical Gram matrix $\frac{1}{n} \sum x_i x_i^\top$ satisfies the Restricted Eigenvalue (RE) condition on support sets of size $s$,
 \\
 (A3): $ \epsilon_i $ has finite second-moment.
\end{assume}

We note that the above assumption follow the standard assumption for high-dimensional robust linear regression as documented in \cite{loh2017statistical,loh2021scale}.

\begin{theorem}
\label{thm_01}
Assume that Assumption \ref{assum_main} holds.  Then, with high probability, for $\lambda \asymp \sqrt{\frac{\log p}{n}}$, the heavy Lasso estimator in \eqref{eq:student_t_lasso} satisfies:
$$
\|\widehat{\beta} - \beta^*\|_2 \lesssim \sqrt{\frac{s \log p}{n}},
\quad
\|\widehat{\beta} - \beta^*\|_1 \lesssim s\sqrt{\frac{ \log p}{n}}
.
$$  
\end{theorem}

The proof of Theorem \ref{thm_01} is provided in Appendix \ref{sc_proof}, and largely follows the general framework outlined in \cite{loh2017statistical,loh2021scale}. 
This theorem delivers non-asymptotic error bounds for the heavy Lasso estimator within a high-dimensional sparse linear regression context. 
It demonstrates that, with high probability, the estimation error in $\ell_2$-norm is of order $\sqrt{\frac{s \log p}{n}}$,
while the $\ell_1$-error scales as $s \sqrt{\frac{\log p}{n}}$, where $s$ denotes the sparsity level of the true coefficient vector $\beta^*$, $p$ is the number of covariates, and $n$ is the sample size. 
These results confirm that the heavy Lasso adapts well to sparsity and achieves consistent estimation even in high-dimensional settings (i.e., $p \gg n$), 
provided that the effective sample size condition $s \log p \ll n$ is met. The theorem also emphasizes the necessity of choosing the regularization parameter $\lambda$ on the scale of $\sqrt{\log p / n}$ to ensure optimal performance. Overall, the result offers a solid theoretical foundation for using heavy Lasso method and these results are similar to those obtained for Lasso using Huber loss as in \cite{loh2017statistical,loh2021scale}.

\section{A data augmentation soft-thresholding algorithm}
\label{sc_algorithm}

Now, we provide details on our development of the data augmentation soft-thresholding algorithm which contain 2 steps:  data augmentation and soft-thresholding steps. Our algorithm is summary in Algorithm~\ref{m_algorithm}. 

The advantage of our new loss function lies in its computational tractability---it requires only one additional step beyond the classical Lasso, making it particularly appealing for practical use.

\subsection{Data augmentation}
The Student-\(t\) distribution can be expressed as a scale mixture of Gaussians:
\[
\epsilon_i \sim t_\nu 
\Longleftrightarrow
\epsilon_i \mid \omega_i \sim \mathcal{N}(0, 1 / \omega_i), 
\quad
\omega_i \sim \text{Gamma}\left(\frac{\nu}{2}, \frac{\nu}{2}\right).
\]
This representation suggests a latent-variable model in which we treat \(\omega_i\) as missing data. The complete-data negative log-likelihood (ignoring constants) is:
\[
L_c(\beta; \omega) = \frac{1}{2n} \sum_{i=1}^n \omega_i (y_i - x_i^\top \beta)^2 + \frac{1}{2n} \sum_{i=1}^n \log \omega_i.
\]
We propose an EM-style algorithm, alternating between:

\textbf{E-step}: Compute the expectation of \(\omega_i\) given current residuals:
\[
\omega_i^{(t)} = \mathbb{E}\left[ \omega_i \mid y_i, \beta^{(t)} \right] =
\frac{\nu + 1}{\nu  + (y_i - x_i^\top \beta^{(t)})^2}.
\]

\textbf{M-step}: Solve a weighted Lasso problem with given weights \( \omega_i^{(t)}\):
\begin{equation}
\beta^{(t+1)} = \arg\min_{\beta \in \mathbb{R}^p}
\left\{
\frac{1}{2n} \sum_{i=1}^n \omega_i^{(t)} (y_i - x_i^\top \beta)^2 + \lambda \|\beta\|_1
\right\}.
\label{eq:weighted_lasso}
\end{equation}

\subsection{Coordinate descent updates using soft-thresholding}

We solve problem \eqref{eq:weighted_lasso} using coordinate descent. Define residuals:
\[
r^{(t)} = y - X \beta^{(t)}.
\]
At each coordinate \(j\), define the partial residual (excluding variable \(j\)):
\[
r_j = r^{(t)} + X_j \beta_j^{(t)}.
\]
The objective restricted to \(\beta_j\) becomes:
\[
\frac{1}{2n} \sum_{i=1}^n w_i (r_{ij} - x_{ij} \beta_j)^2 + \lambda |\beta_j|.
\]
This is minimized by the soft-thresholding operator, \citep{hastie2009elements},:
\[
z_j := \sum_{i=1}^n w_i x_{ij} r_{ij}, \quad
A_j := \sum_{i=1}^n w_i x_{ij}^2, \quad
\beta_j \leftarrow \frac{1}{A_j} 
\mathcal{S}(z_j, \lambda),
\]
where 
$$
\mathcal{S}
(z, \lambda) = \text{sign}(z) \cdot \max(|z| - \lambda, 0)
.
$$

\begin{algorithm}[H]
\caption{Data augmentation soft-thresholding algorithm for heavy Lasso}
\begin{algorithmic}[1]
\STATE \textbf{Input:} Data \(X \in \mathbb{R}^{n \times p}\), \(y \in \mathbb{R}^n\), regularization \(\lambda\), degrees of freedom \(\nu\), max iterations \(T\)
\STATE Initialize \(\beta^{(0)} = 0\)
\FOR{$t = 1$ to $T$}
    \STATE Compute residuals: \(r_i = y_i - x_i^\top \beta^{(t)}\)
    \STATE Compute weights: \(w_i = \frac{\nu + 1}{\nu  + r_i^2}\)
    \FOR{$j = 1$ to $p$}
        \STATE Partial residual: \(r_j = y - X \beta^{(t)} + X_j \beta_j^{(t)}\)
        \STATE Compute: \(z_j = \sum_{i=1}^n w_i x_{ij} r_{ij}\), \(A_j = \sum_{i=1}^n w_i x_{ij}^2\)
        \STATE Update: \(\beta_j^{(t+1)} = \frac{1}{A_j} \mathcal{S}(z_j, \lambda)\)
    \ENDFOR
\ENDFOR
\STATE \textbf{Output:} Final estimate \(\hat{\beta} = \beta^{(T)}\)
\end{algorithmic}
\label{m_algorithm}
\end{algorithm}

\subsection{Tuning $\lambda$ via information criteria}

To select the regularization parameter $\lambda$ in the heavy Lasso method, we adopt a model selection approach based on information criteria---specifically, the Bayesian Information Criterion (BIC). 
For each candidate $\lambda$, the fitted model produces a set of coefficients $\hat{\boldsymbol{\beta}}_\lambda$, and we compute the residual sum of squares (RSS) as
$$
\text{RSS}(\lambda) = \sum_{i=1}^n \left(y_i - \mathbf{x}_i^\top \hat{\boldsymbol{\beta}}_\lambda \right)^2.
$$
Let $df_\lambda = \|\hat{\boldsymbol{\beta}}_\lambda\|_0$ denote the number of non-zero estimated coefficients, which serves as an estimate of the degrees of freedom. The BIC for each model are then calculated as
$$
\text{BIC}(\lambda) = n \log\left( \frac{\text{RSS}(\lambda)}{n} \right) + \log(n) \cdot df_\lambda,
$$
where $n$ is the number of observations. The optimal value of $\lambda$ is selected as the one minimizing BIC.  This procedure provides a robust, data-driven criterion for regularization in heavy-tailed regression settings. We have also tested the Akaike Information Criterion (AIC) but the results are not good as BIC. AIC tends to select models with better predictive performance by allowing more flexibility, while BIC imposes a stronger penalty on model complexity, typically leading to more parsimonious models.

Our method is implemented in the \texttt{R} package \texttt{heavylasso} available on Github: \url{https://github.com/tienmt/heavylasso} .

\section{Simulation studies}
\label{sc_simulations}
The main objective of this simulation study is to assess the performance of Lasso estimators under various robust loss functions. Therefore, we restrict our comparison to methods that employ the Lasso penalty. Alternative regularization approaches or penalty types are not considered in this study.

\subsection{Compared methods and setup}
We consider the following four alternative Lasso methods:
\begin{itemize}
    \item Classical Lasso using the squared loss function, implemented in the \texttt{R} package \texttt{glmnet} \citep{glmnetpackage}.
    
    \item LAD Lasso using the $\ell_1$ loss function, available via the \texttt{R} package \texttt{hqreg} \citep{yi2017semismooth}.
    
    \item Huber Lasso using the Huber loss function, also implemented in the package \texttt{hqreg}.
    
    \item Rank-based Lasso using a rank-based loss function, provided by the \texttt{R} package \texttt{TFRE} \citep{wang2020tuning}.
\end{itemize}

\noindent We generate predictors $X_i \sim N(0, \Sigma)$, and consider two covariance structures for $\Sigma$:
\begin{itemize}
    \item Independent predictors: $\Sigma = \mathbb{I}_p$;
    \item Correlated predictors: $\Sigma_{ij} = \rho_X^{|i-j|}$ for all $i, j$.
\end{itemize}
For a given sparsity level $s$, the first half of the non-zero entries in $\beta_0$ are set to 1, and the second half to –1. The responses $y_i$ are then generated according to the linear model \eqref{eq_linear_model}, under the following noise settings:
\begin{itemize}
    \item Gaussian noise, $ \epsilon_i \sim \mathcal{N} (0,1) $. This serves as a baseline to assess how various robust methods perform under ideal (light-tailed) conditions.
    
    \item Gaussian with outlier noise. $ \epsilon_i \sim \mathcal{N} (0,1) $ but 30\% of the observed responses are further contaminated by outliers. This setting evaluates robustness to contamination.
    
    \item Gaussian with large variance. $ \epsilon_i \sim \mathcal{N} (0,3^2) $.
    This setting introduces moderate heavy-tailed behavior through increased variance.
    
    \item Student noise. $ \epsilon_i \sim t_3 $. This case represents heavy-tailed noise with finite variance.
    
    \item Cauchy noise. $ \epsilon_i \sim Cauchy $. This represents a more extreme heavy-tailed setting with infinite variance. 
\end{itemize}

To evaluate estimation accuracy, we consider the $\ell_1$ norm $ \| \widehat{\beta} - \beta_0 \|_1$ and the squared $\ell_2 $ norm $\| \widehat{\beta} - \beta_0 \|_2^2 $, 
where $\widehat{\beta}$ denotes the estimator obtained by each method. We also report the error in the linear predictor,
defined as $n^{-1} \| X^\top (\widehat{\beta} - \beta_0) \|_2^2 \, $. For assessing prediction accuracy, we use the following prediction error metric:
$$
\text{prediction} := 
\frac{1}{n_{\text{test}}} \sum_{i=1}^{n_{\text{test}}} 
\left( y_{{\text{test}},i} - 
X_{{\text{test}},i}^\top \widehat{\beta}
\right)^2,
$$
where $(X_{\text{test}}, y_{\text{test}})$ are independent test data generated from the same model as the training data using the true parameter $\beta_0$. We fix the test sample size at $n_{\text{test}} = 50$ for all simulation settings.

Each simulation setting is repeated 50 times, and we report the mean and standard deviation of the results. The outcomes are presented in Tables \ref{tb_low_dim}, \ref{tb_low_dim_rX}, \ref{tb_high_dim}, and \ref{tb_high_dim_rX}.
For the Lasso methods with squared loss, $\ell_1$ loss, and Huber loss, tuning parameters are selected via 5-fold cross-validation. The rank-based Lasso is a tuning-free method and does not require parameter selection. Our proposed Heavy Lasso method is tuned using the AIC.

\subsection{Results}

\subsubsection{Results with Gaussian type noise}

Firstly, we observe that both our proposed Heavy Lasso method and the classical Lasso, which utilizes the squared loss, consistently demonstrate superior performance when the noise distribution is Gaussian or contaminated by Gaussian noise. Notably, the Heavy Lasso occasionally surpasses the classical Lasso in predictive accuracy, as evidenced in Tables \ref{tb_low_dim} and \ref{tb_high_dim}.

As anticipated, our Heavy Lasso significantly outperforms variants of the Lasso that employ 
$ \ell_1 $	
  loss or rank-based loss functions, which tend to be less effective in these settings.

When compared to the Lasso variant incorporating Huber loss, our Heavy Lasso frequently exhibits very similar performance, even in cases where it is not the absolute best. Moreover, under scenarios involving Gaussian noise with large variance, the Heavy Lasso consistently achieves lower prediction errors than the Huber loss-based Lasso, underscoring its robustness and adaptability across varying noise structures.

\subsubsection{Results with heavy tailed noise}

In the presence of heavy-tailed noise, specifically when the errors follow a $t_3$ distribution, our Heavy Lasso method frequently emerges as the best-performing approach, achieving the lowest prediction error across various scenarios. Regarding estimation accuracy, it maintains superior performance even in higher-dimensional settings, as demonstrated in Tables \ref{tb_high_dim} and \ref{tb_high_dim_rX}. However, in cases with smaller sample sizes and lower dimensionality, its performance becomes comparable to that of the Lasso employing the Huber loss. Importantly, our method consistently outperforms alternative loss-based approaches, underscoring its robustness and effectiveness in modeling data with heavy-tailed noise.

\subsubsection{Results with extremely heavy tails}

Consider the particularly challenging scenario where the noise follows a Cauchy distribution, which lacks a finite variance. In this setting, all competing methods struggle to achieve satisfactory prediction accuracy. Although our Heavy Lasso method also underperforms with small sample sizes and low-dimensional data, it demonstrates a remarkable advantage in higher-dimensional contexts, as illustrated in Tables \ref{tb_high_dim} and \ref{tb_high_dim_rX}. Specifically, our method is the only approach capable of attaining substantially low estimation error under these conditions. This result is both striking and insightful, highlighting a scenario where our Heavy Lasso outperforms the Huber loss, thereby underscoring its robustness and effectiveness in handling extreme heavy-tailed noise.

\begin{table}[!ht]
\centering
\caption{Simulation results with $ p = 120, s^* = 20, n = 100 $.}
	\begin{tabular}{ l l | cccc }
		\hline \hline
Noise & Method  
& $ \| \widehat{\beta} -\beta_0 \|_2^2 $ 
& $ \| \widehat{\beta} -\beta_0 \|_1 $   
& $ \frac{1}{n}\| X^\top\! (\widehat{\beta} -\beta_0) \|_2^2 $ 
& prediction
\\
\hline \hline
$ \mathcal{N} (0,1) $  & proposed loss
& \textbf{1.39} (0.56) & 7.16 (1.40) & \textbf{0.61} (0.14) & \textbf{2.37} (0.67)
\\
& squared loss 
& 1.59 (0.74) & \textbf{5.99} (1.41) &  0.74 (0.25) &  2.47 (0.75)
\\
& $\ell_1$ loss 
& 1.94 (0.78) & 7.56 (1.61) & 0.92 (0.25) & 2.83 (0.86) 
\\
& Huber loss
& 1.65 (0.71) & 6.50 (1.50) & 0.77 (0.24) & 2.55 (0.75) 
\\
& rank-based loss
& 18.8  (0.80) & 19.2  (0.56) & 18.2  (2.27) &  19.6  (4.19)
\\
		\hline
		\hline	
$ \mathcal{N} (0,1) $, & proposed loss
& \textbf{1.48} (0.48) & 7.34 (1.24) & \textbf{0.63} (0.12) &  \textbf{2.39} (0.60)
\\
 outlier 30\% & squared loss 
& 1.56 (0.61) & \textbf{6.00} (1.42) & 0.71 (0.20) &  2.47 (0.78)
\\
& $\ell_1$ loss 
& 2.22 (1.00) & 8.00 (1.71) & 0.97 (0.34) & 3.11 (1.11) 
\\
& Huber loss
& 1.74 (0.73) & 6.73 (1.54) & 0.79 (0.25) &  2.67 (0.84)
\\
& rank-based loss
& 19.1  (0.58) & 19.4  (0.40) & 18.0  (2.11) &  20.0  (4.05)
\\
		\hline
		\hline	
$ \mathcal{N} (0,3) $  & proposed loss
& 14.6 (2.88) & 23.9  (3.63) & \textbf{7.73} (2.45) & 24.6  (5.55)  
\\
& squared loss 
& \textbf{13.5}  (3.76) & \textbf{16.4}  (2.30) & 10.2  (4.47) &  \textbf{23.2}  (6.70)
\\
& $\ell_1$ loss 
& 15.8  (3.69) & 18.4  (2.17) & 12.6  (5.25) &  25.9  (6.75)
\\
& Huber loss
& 15.2  (3.59) & 18.0  (2.12) & 11.7  (4.69) & 25.2  (6.75) 
\\
& rank-based loss
& 19.4  (0.47) & 19.7  (0.31) & 19.1  (2.73) & 29.4  (6.14) 
\\
		\hline
		\hline	
$ t_3 $ & proposed loss
& \textbf{3.85} (3.66) & 10.7  (3.27) & \textbf{2.01} (2.77) &  \textbf{6.24} (3.10)
\\
& squared loss 
& 6.06 (5.13) & 10.7  (3.93) & 3.74 (4.57) &  8.51 (4.13)
\\
& $\ell_1$ loss 
& 4.63 (3.35) & 10.5  (3.37) & 2.20 (1.77) & 7.12 (3.52) 
\\
& Huber loss
& 4.41 (3.52) & \textbf{9.67} (3.67) & 2.11 (1.74) & 7.11 (3.97) 
\\
& rank-based loss
& 19.0  (0.71) & 19.4  (0.49) &  18.3  (2.65) & 21.7  (4.69) 
\\
		\hline
		\hline	
$ Cauchy $ & proposed loss
& 19.6  (0.98) & 19.7  (0.51) & 19.0 (3.30) & 466 (865) 
\\
& squared loss 
& 19.9  (0.08) & 19.9  (0.04) & 19.9  (2.85) &  466 (864)
\\
& $\ell_1$ loss 
& 18.9  (2.31) & 19.3  (1.41) & 18.1  (4.17) & 466 (866) 
\\
& Huber loss
& 19.0  (1.95) & 19.4  (1.14) & 18.3 (3.97) &  466 (866)
\\
& rank-based loss
& 19.4  (0.50) & 19.6 (0.35) & 18.8  (2.47) &  466 (864)
\\
		\hline
		\hline	
\end{tabular}
\label{tb_low_dim}
\end{table}

\begin{table}[!ht]
\centering
\caption{Simulation results for correlated $X, \rho_X = 0.5 $ and with $ p = 120, s^* = 20, n = 100 $.}
	\begin{tabular}{ l l | cccc }
		\hline \hline
Noise & Method  
& $ \| \widehat{\beta} -\beta_0 \|_2^2 $ 
& $ \| \widehat{\beta} -\beta_0 \|_1 $   
& $ n^{-1}\| X^\top\! (\widehat{\beta} -\beta_0) \|_2^2 $ 
& prediction
\\
\hline \hline
$ \mathcal{N} (0,1) $  & proposed loss
& 1.20 (0.33) & 6.34 (0.84) & \textbf{0.51} (0.09) &  \textbf{1.89} (0.40)
\\
& squared loss 
& \textbf{1.02} (0.37) & \textbf{4.07} (0.85) & 0.55 (0.18) &  1.90 (0.43)
\\
& $\ell_1$ loss 
& 1.52 (0.54) & 5.41 (0.90) & 0.73 (0.23) &  2.17 (0.54)
\\
& Huber loss
& 1.13 (0.42) & 4.36 (0.84) & 0.59 (0.16) & 1.95 (0.48) 
\\
& rank-based loss
& 12.1 (2.36) & 14.1  (1.95) & 16.3  (5.26) &  23.0  (8.99)
\\
		\hline
		\hline	
$ \mathcal{N} (0,1) $ & proposed loss
& 1.22 (0.31) & 6.35 (0.82) & \textbf{0.52} (0.09) &  \textbf{1.96} (0.54)
\\
outlier 30\%  & squared loss 
& \textbf{1.09} (0.37) & \textbf{4.12} (0.68) & 0.60 (0.16) &  2.01 (0.54) 
\\
& $\ell_1$ loss 
& 1.77 (0.75) & 5.82 (1.12) & 0.83 (0.30) & 2.44 (0.70) 
\\
& Huber loss
& 1.32 (0.60) & 4.56 (0.91) & 0.66 (0.21) & 2.12 (0.55) 
\\
& rank-based loss
& 11.4  (2.62) & 13.6  (2.08) & 14.9  (4.84) &  21.9  (8.55)
\\
		\hline
		\hline	
$ \mathcal{N} (0,3) $  & proposed loss
& 11.6  (2.79) &  19.4  (4.28) & 6.96 (3.32) & 20.1  (6.74) 
\\
& squared loss 
& \textbf{7.40} (1.76) & \textbf{11.0}  (1.69) & \textbf{5.50} (1.77) &  \textbf{17.1}  (4.43)
\\
& $\ell_1$ loss 
& 9.52 (1.89) & 13.3  (1.80) & 7.40 (2.74) & 19.5  (4.20) 
\\
& Huber loss
& 9.14 (1.76) & 13.0  (2.00) & 6.81 (2.59) &  18.8 (4.14)
\\
& rank-based loss
& 14.6  (1.78) & 16.0  (1.43) & 23.4  (4.25) & 37.5 (10.3) 
\\
		\hline
		\hline	
$ t_3 $  & proposed loss
& 2.91 (1.06) & 9.05 (2.33) & \textbf{1.23} (0.36) &  \textbf{5.26} (2.35)
\\
& squared loss 
& 3.79 (2.12) & 7.53 (1.91) & 2.17 (1.74) & 6.61 (4.01) 
\\
& $\ell_1$ loss 
& 3.26 (1.29) & 7.56 (1.48) & 1.62 (0.85) & 5.70 (2.25) 
\\
& Huber loss
& \textbf{2.77} (1.09) & \textbf{6.58} (1.28) & 1.39 (0.53) & 5.30 (2.31) 
\\
& rank-based loss
& 12.4  (2.04) & 14.3  (1.61) & 15.8  (3.78) & 23.9  (7.30) 
\\
		\hline
		\hline	
$ Cauchy $  & proposed loss
& 17.0  (4.46) & 18.0  (3.26) & 36.3 (16.8) & 245 (529) 
\\
& squared loss 
& 18.2  (3.23) & 18.7  (2.39) & 40.6 (14.4) &  251 (532) 
\\
& $\ell_1$ loss 
& 12.2  (4.93) & 14.3  (3.69) & 18.8 (16.3) &  233 (538)
\\
& Huber loss
& 12.1  (5.13) & 14.2  (3.91) & 19.1 (16.7) & 233 (536) 
\\
& rank-based loss
& 15.5  (1.99) & 16.7  (1.62) & 26.8  (6.91) &  240 (534)
\\
		\hline
		\hline	
\end{tabular}
\label{tb_low_dim_rX}
\end{table}

\begin{table}[!ht]
\centering
\caption{Simulation results with $ p = 500, s^* = 20, n = 300 $.}
	\begin{tabular}{ l l | cccc }
		\hline \hline
Noise & Method  
& $ \| \widehat{\beta} -\beta_0 \|_2^2 $ 
& $ \| \widehat{\beta} -\beta_0 \|_1 $   
& $ \frac{1}{n}\| X^\top\! (\widehat{\beta} -\beta_0) \|_2^2 $ 
& prediction
\\
\hline \hline
$ \mathcal{N} (0,1) $  & proposed loss
& 0.64 (0.18) & 3.55 (0.52) & 0.50 (0.12) & 1.59 (0.37)
\\
& squared loss 
& \textbf{0.52} (0.14) & \textbf{3.43} (0.44) & \textbf{0.42} (0.10) & \textbf{1.51} (0.38)
\\
& $\ell_1$ loss 
& 0.70 (0.17) & 4.62 (0.65) & 0.53 (0.12) & 1.66 (0.39)
\\
& Huber loss
& 0.59 (0.15) & 3.89 (0.55) & 0.46 (0.11) & 1.57 (0.37)
\\
& rank-based loss
& 13.1 (3.07) & 15.7 (2.23) & 12.2 (2.47) & 13.8  (4.00)
\\
		\hline
		\hline	
$ \mathcal{N} (0,1) $, & proposed loss
& 0.61 (0.15) & \textbf{3.50} (0.48) & 0.48 (0.11) & 1.65 (0.37)
\\
 outlier 30\% & squared loss 
& \textbf{0.55} (0.15) & 3.61 (0.55) & \textbf{0.42} (0.11) & \textbf{1.51} (0.37)
\\
& $\ell_1$ loss 
& 0.76 (0.23) & 4.93 (0.80) & 0.55 (0.17) & 1.74 (0.42)
\\
& Huber loss
& 0.64 (0.18) & 4.10 (0.67) & 0.47 (0.13) & 1.61 (0.39)
\\
& rank-based loss
& 13.4  (2.71) & 15.9  (1.86) & 12.2 (2.14) & 14.1 (3.74)
\\
		\hline
		\hline	
$ \mathcal{N} (0,3) $  & proposed loss
&  \textbf{3.93} (1.00) & 13.7 (2.03) & \textbf{2.84} (0.55) & \textbf{12.7}  (3.29) 
\\
& squared loss 
& 4.71 (1.34) & \textbf{10.3}  (1.42) & 3.72 (0.96) & 13.3  (3.28)
\\
& $\ell_1$ loss 
& 7.12 (2.04) & 14.5 (2.13) & 5.28 (1.58) & 15.8 (4.74)
\\
& Huber loss
& 6.55 (1.76) & 13.3  (1.89) & 4.90 (1.31) & 15.0  (4.06)
\\
& rank-based loss
& 17.6 (1.08) & 18.6 (0.69) & 17.0 (1.19) & 25.7 (6.40)
\\
		\hline
		\hline	
$ t_3 $ & proposed loss
& \textbf{1.04} (0.25) & \textbf{4.99} (0.68) & \textbf{0.77} (0.17) & \textbf{4.17} (1.92)
\\
& squared loss 
& 1.89 (1.24) & 6.19 (1.47) & 1.50 (0.99) & 5.13 (2.60)
\\
& $\ell_1$ loss 
& 1.18 (0.32) & 5.71 (0.88) & 0.86 (0.24) & 4.37 (2.22)
\\
& Huber loss
& 1.08 (0.29) & 5.07 (0.69) & 0.82 (0.23) & 4.24 (2.16)
\\
& rank-based loss
& 15.2 (1.80) & 17.1 (1.17) & 13.9 (1.45) & 18.7 (4.88)
\\
		\hline
		\hline	
$ Cauchy $ & proposed loss
& \textbf{9.60} (4.81) & \textbf{13.4}  (4.15) & \textbf{7.95} (3.95) & 1516 (7386)
\\
& squared loss 
& 19.7 (1.54) & 19.8 (1.02) & 19.5 (2.26) & 1527 (7393)
\\
& $\ell_1$ loss 
& 11.7  (6.70) & 14.6 (4.52) & 10.8 (6.99) & 1515 (7374)
\\
& Huber loss
& 11.4 (6.79) &  14.3 (4.74) & 10.6  (6.94) & 1516 (7375)
\\
& rank-based loss
& 18.0 (1.27) & 18.8 (0.82) & 17.3 (1.26) & 1526 (7399)
\\
		\hline
		\hline	
\end{tabular}
\label{tb_high_dim}
\end{table}

\begin{table}[!ht]
\centering
\caption{Simulation results correlated $X, \rho_X = 0.5$  with $ p = 500, s^* = 20, n = 300 $.}
	\begin{tabular}{ l l | cccc }
		\hline \hline
Noise & Method  
& $ \| \widehat{\beta} -\beta_0 \|_2^2 $ 
& $ \| \widehat{\beta} -\beta_0 \|_1 $   
& $ \frac{1}{n}\| X^\top\! (\widehat{\beta} -\beta_0) \|_2^2 $ 
& prediction
\\
\hline \hline
$ \mathcal{N} (0,1) $  & proposed loss
& \textbf{0.37} (0.11) & 2.33 (0.34) &  \textbf{0.30} (0.10) & 1.36 (0.31)
\\
& squared loss 
& \textbf{0.37} (0.11) & \textbf{2.24} (0.32) & 0.31 (0.07) & \textbf{1.33} (0.27)
\\
& $\ell_1$ loss 
& 0.55 (0.18) & 3.03 (0.59) & 0.42 (0.11) & 1.48 (0.32)
\\
& Huber loss
& 0.43 (0.13) & 2.44 (0.38) & 0.34 (0.08) & 1.37 (0.28)
\\
& rank-based loss
& 1.26 (0.44) & 3.70 (0.63) & 1.15 (0.30) & 2.26 (0.66)
\\
		\hline
		\hline	
$ \mathcal{N} (0,1) $, & proposed loss
& 0.40 (0.12) & 2.40 (0.38) & 0.32 (0.09) & 1.37 (0.32)
\\
 outlier 30\% & squared loss 
& \textbf{0.37} (0.12) & \textbf{2.30} (0.33) & \textbf{0.30} (0.09) & \textbf{1.35} (0.27)
\\
& $\ell_1$ loss 
& 0.51 (0.15) & 3.04 (0.47) & 0.39 (0.10) & 1.46 (0.28)
\\
& Huber loss
& 0.42 (0.14) & 2.48 (0.38) & 0.34 (0.09) & 1.38 (0.27)
\\
& rank-based loss
& 1.36 (0.48) & 3.90 (0.69) & 1.23 (0.31) &  2.45 (0.68)
\\
		\hline
		\hline	
$ \mathcal{N} (0,3) $  & proposed loss
& 3.24 (0.80) & 12.1 (1.87) & \textbf{2.34} (0.44) &  \textbf{12.3}  (2.53)
\\
& squared loss 
& \textbf{3.13} (0.79) & \textbf{6.77} (0.74) & 2.72 (0.72) & 12.5 (2.55)
\\
& $\ell_1$ loss 
& 4.23 (0.97) & 8.51 (1.04) & 3.61 (1.03) & 13.6 (2.95)
\\
& Huber loss
& 3.95 (0.98) & 7.93 (0.97) & 3.43 (1.04) & 13.4  (2.80)
\\
& rank-based loss
& 6.55 (1.29) & 9.62 (1.15) & 7.74 (1.70) & 18.1 (4.00)
\\
		\hline
		\hline	
$ t_3 $ & proposed loss
& \textbf{0.64} (0.23) & 3.53 (0.61) & \textbf{0.45} (0.13) & \textbf{3.21} (0.92)
\\
& squared loss 
& 1.30 (0.86) & 4.03 (1.15) & 1.22 (1.29) & 3.87 (3.02)
\\
& $\ell_1$ loss 
& 0.86 (0.30) & 3.66 (0.65) & 0.64 (0.23) & 3.27 (2.91)
\\
& Huber loss
& 0.75 (0.24) & \textbf{3.16} (0.54) & 0.60 (0.16) &  3.23 (2.97)
\\
& rank-based loss
& 2.24 (0.74) & 5.03 (0.90) & 2.08 (0.64) & 4.82 (2.84)
\\
		\hline
		\hline	
$ Cauchy $ & proposed loss
& \textbf{1.26} (0.54) & \textbf{4.20} (0.82) & \textbf{1.02} (0.35) & 12301 (77648)
\\
& squared loss 
& 18.9  (3.41) & 19.1  (2.63) & 45.3 (11.2) & 12366 (77805)
\\
& $\ell_1$ loss 
& 8.31 (5.83) & 10.8 (4.83) & 14.8 (15.4) & 12326 (77723)
\\
& Huber loss
& 8.45 (5.82) & 10.9  (4.82) & 15.3 (15.4) & 12327 (77722)
\\
& rank-based loss
& 5.86 (1.61) & 8.89 (1.51) & 7.02 (2.50) & 12313 (77696)
\\
		\hline
		\hline	
\end{tabular}
\label{tb_high_dim_rX}
\end{table}

\section{Application: predicting protein expression based on gene expression data}
\label{sc_real_data}

We use a pre-processed version of the NCI-60 cancer cell line panel, as described in \cite{reinhold2012cellminer}. The dataset consists of gene and protein expression measurements for 59 cancer cell lines ($n=59$), after excluding one observation with missing gene expression data. This data is available from the \texttt{R} package \texttt{robustHD} \citep{robusthdpackage}. 
The gene expression data comprise a matrix with 22,283 features. The protein expression data are stored in a matrix with 162 features, derived from reverse-phase protein lysate arrays and $\log_2$-transformed. This dataset provides a rich framework for modeling protein expression from high-dimensional gene expression profiles in a biological context. Following \citep{robusthdpackage}, we select protein 92 as the response variable and reduce the dimensionality of the covariate matrix by retaining the 150 gene expression features most correlated with it.

We randomly leave out 9 samples as test data and train all considered methods on the remaining 50 samples. 
This process is repeated 100 times, and the average prediction error on the test data is reported in Table \ref{tb_realdata}. 
The results show that the rank-based Lasso achieves the lowest prediction error. Our proposed heavy Lasso method is comparable to the Huber loss and absolute loss approaches. In contrast, the classical Lasso clearly performs poorly on this dataset.

\begin{table}[!h]
	\centering
	\caption{Prediction errors for real data application.}
	\begin{tabular}{  ccccc }
		\hline \hline
proposed loss & squared loss & $\ell_1$ loss & Huber loss & rank-based loss 	
		\\ 		\hline
7.86 (2.28) & 11.4 (3.65) & 7.83 (2.45) & 7.81 (2.45) & 7.55 (2.34) 
		\\
		\hline
		\hline
	\end{tabular}
	\label{tb_realdata}
\end{table}

\section{Conclusion}
\label{sc_conclusion}

In this paper, we have introduced the Heavy Lasso, a novel robust penalized regression method specifically designed for high-dimensional linear models subject to heavy-tailed noise and potential outliers. Unlike the classical Lasso, which relies on the squared loss and implicitly assumes Gaussian-like errors, our method incorporates a robust loss function derived from the Student-t distribution. This construction allows Heavy Lasso to maintain the desirable quadratic behavior around small residuals while adaptively downweighting the impact of extreme deviations, thus offering enhanced robustness against heavy-tailed contamination and outliers frequently encountered in many modern applications.

Our proposed approach achieves a compelling balance between statistical robustness and computational efficiency. By leveraging a data augmentation framework, the Heavy Lasso estimator can be computed using a soft-thresholding iterative algorithm that integrates seamlessly with standard Lasso optimization routines. This avoids the often significant computational overhead associated with non-convex or complex robust loss functions, making our method both practical and scalable for large-scale problems.

The theoretical analysis presented here further strengthens the appeal of Heavy Lasso by establishing non-asymptotic error bounds under a localized convexity framework. These results hold under minimal moment assumptions on the noise distribution, thereby accommodating a wide range of heavy-tailed settings beyond the classical sub-Gaussian paradigm. The theoretical guarantees mirror those available for other robust M-estimators such as the Huber loss, confirming that our Student-t inspired loss shares similarly favorable statistical properties.

Extensive simulation studies demonstrate that Heavy Lasso consistently competes with or outperforms alternative state-of-the-art robust penalized regression methods, particularly in scenarios involving heavy-tailed noise or contaminated data. Notably, our method shows strong advantages in estimation accuracy and prediction performance when noise distributions exhibit heavy tails, while maintaining comparable performance to classical Lasso under Gaussian noise conditions.

Several promising research directions emerge from this work. One avenue is to extend our loss function to incorporate non-convex penalties, such as SCAD or MCP. Another extension involves adapting the framework to handle structured sparsity, such as group or hierarchical variable selection, which is particularly relevant in applications like genomics or image analysis. Developing valid and efficient confidence intervals under heavy-tailed settings is also an important goal, requiring careful consideration of the theoretical properties of the estimator \citep{loh2021scale}.
A particularly important direction for future research is the extension of our methodology to a Bayesian framework.
Recent advances in generalized Bayesian inference provide a solid foundation for this, especially under model misspecification  \citep{mai2025concentration, mai2024concentration, mai2025hightobit}. Leveraging these developments may lead to robust, flexible, and theoretically grounded Bayesian analogs of our current approach.

\subsection*{Acknowledgments}
The findings, interpretations, and conclusions expressed in this paper are entirely those of the author and do not reflect the views or positions of the Norwegian Institute of Public Health in any forms.

\subsection*{Conflicts of interest/Competing interests}
The author declares no potential conflict of interests.

\clearpage
\appendix
\section{Proofs}
\label{sc_proof}

Let $$
L(\beta) = \frac{\nu + 1}{2n} \sum_{i=1}^n \log\left(1 + \frac{(y_i - x_i^\top \beta)^2}{\nu }\right).
$$

\begin{proof}[\bf Proof of Theorem \ref{thm_01}]

Let $\Delta := \widehat{\beta} - \beta^*$ be the estimation error.

We proceed in three steps:

\noindent Step 1: First-order optimality and decomposition.
\\
The optimality of $\widehat{\beta}$ implies:
$$
L(\widehat{\beta}) + \lambda \|\widehat{\beta}\|_1 \le L(\beta^*) + \lambda \|\beta^*\|_1.
$$
Rewriting:
$$
L(\beta^* + \Delta) - L(\beta^*) \le \lambda \left( \|\beta^*\|_1 - \|\beta^* + \Delta\|_1 \right).
$$
Let $S := \operatorname{supp}(\beta^*)$. Then, see for example \cite{loh2021scale},
$$
\|\beta^*\|_1 - \|\beta^* + \Delta\|_1 
\leq
\|\Delta_S\|_1 - \|\Delta_{S^c}\|_1.
$$
Thus:
$$
L(\beta^* + \Delta) - L(\beta^*) \le \lambda (\|\Delta_S\|_1 - \|\Delta_{S^c}\|_1).
$$

\noindent Step 2: Local Restricted Strong Convexity (LRSC)
\\
The key is a localized strong convexity inequality. From Lemma \ref{lm_GRSC}, we have that
\begin{equation}
    \label{eq_lrsc_01}
L(\beta^* + \Delta) - L(\beta^*) \ge \langle \nabla L(\beta^*), \Delta \rangle + \frac{\kappa}{2} \|\Delta\|_2^2.    
\end{equation}
for some $\kappa > 0$ when $\|\Delta\|_2$ is not too large.

\noindent Step 3: Control stochastic term
\\
From Lemma \ref{thm_002}, for $\lambda \asymp \sqrt{\frac{\log p}{n}}$, with probability at least $1 - c/p$, we have that
$$
\|\nabla L(\beta^*)\|_\infty \le \lambda/2.
$$
Thus:
\begin{equation}
    \label{eq_controlstochastic_02}
| \langle \nabla L(\beta^*), \Delta \rangle |
\leq
\|\nabla L(\beta^*)\|_\infty \|\Delta\|_1 \le \frac{\lambda}{2} \|\Delta\|_1
.
\end{equation}
Using \eqref{eq_lrsc_01} and \eqref{eq_controlstochastic_02}, we get:
$$
\frac{\kappa}{2} \|\Delta\|_2^2 \le \lambda \left( \|\Delta_S\|_1 - \|\Delta_{S^c}\|_1 \right) + \frac{\lambda}{2} \|\Delta\|_1.
$$
Group terms:
\begin{equation}
    \label{eq_cone_condition}
\frac{\kappa}{2} \|\Delta\|_2^2 
\le 
\frac{3\lambda}{2} \|\Delta_S\|_1 - \frac{\lambda}{2} \|\Delta_{S^c}\|_1
.
\end{equation}
Apply the standard inequality $\|\Delta_S\|_1 \le \sqrt{s} \|\Delta\|_2$, and discard the negative term:
$$
\frac{\kappa}{2} \|\Delta\|_2^2 \le \frac{3\lambda}{2} \sqrt{s} \|\Delta\|_2 \Rightarrow \|\Delta\|_2 \le \frac{3\lambda \sqrt{s}}{\kappa}.
$$

\noindent Therefore, with high probability, we get that
$$
\|\widehat{\beta} - \beta^*\|_2 = \|\Delta\|_2 \le \frac{3\lambda \sqrt{s}}{\kappa} \lesssim \sqrt{\frac{s \log p}{n}}
.
$$
Finally, the inequality in \eqref{eq_cone_condition} implies that 
$$  
\|\Delta_{S^c}\|_1 \leq  3\|\Delta_S\|_1 
$$ 
and that
\begin{equation*}
\|\widehat{\beta} - \beta^*\|_1
=
 \|\Delta_{S}\|_1 + \|\Delta_{S^c}\|_1
 \leq
 4\|\Delta_{S}\|_1
  \leq
 4 \sqrt{s} \|\Delta_{S}\|_2
  \leq
 4 \sqrt{s} \|\widehat{\beta} - \beta^*\|_2
 ,
\end{equation*}
this gives the desired $\ell_1$-bound.
The proof is completed.
    
\end{proof}

\begin{lemma}
\label{thm_002}
Assume that Assumption \ref{assum_main} holds.  Then there exist universal constants $c_1, c_2 > 0$, depending only on $\nu, , K$, such that for any $\delta \in (0, 1)$, with probability at least $1 - \delta$,
$$
\left\|\nabla L(\beta^*)\right\|_\infty \le c_1 \sqrt{\frac{\log(p/\delta)}{n}}.
$$  
\end{lemma}

\begin{proof}[\bf Proof of Lemma \ref{thm_002}]

 Step 1: Gradient expression.
\\
Recall that
$
L(\beta) = \frac{\nu + 1}{2n} \sum_{i=1}^n \log\left(1 + \frac{(y_i - x_i^\top \beta)^2}{\nu }\right).
$
Differentiating with respect to $\beta$, we get:
$$
\nabla L(\beta) = -\frac{\nu + 1}{n} \sum_{i=1}^n \frac{(y_i - x_i^\top \beta)}{\nu  + (y_i - x_i^\top \beta)^2} x_i.
$$
Evaluated at the true parameter $\beta = \beta^*$, this becomes:
$$
\nabla L(\beta^*) = -\frac{\nu + 1}{n} \sum_{i=1}^n w_i x_i, \quad \text{where } w_i := \frac{\epsilon_i}{\nu  + \epsilon_i^2}.
$$
Thus, the $j$-th coordinate of the gradient is:

$$
\left[\nabla L(\beta^*)\right]_j = -\frac{\nu + 1}{n} \sum_{i=1}^n w_i x_{ij}.
$$
\noindent Step 2: Boundedness of $w_i$.
\\
We have:
$$
|w_i| = \left| \frac{\epsilon_i}{\nu  + \epsilon_i^2} \right| \le \max_{t \in \mathbb{R}} \frac{|t|}{\nu  + t^2}.
$$
This function is maximized at $t = \sqrt{\nu} $, so:
$$
\max_t \frac{|t|}{\nu  + t^2} = \frac{\sqrt{\nu} }{2 \nu } = \frac{1}{2  \sqrt{\nu}}.
$$
Hence,
$$
|w_i| \le \frac{1}{2  \sqrt{\nu}} := B.
$$

\noindent Step 3: Concentration inequality
\\
Define:
$$
Z_{ij} := w_i x_{ij}.
$$
Then each 
$$
|Z_{ij}| \le K  B := M
,
$$ since $|x_{ij}| \le K$ and $|w_i| \le B$.

We also note that $\mathbb{E}[w_i] = 0$, so $\mathbb{E}[Z_{ij}] = 0$.
Thus, we can now apply Bernstein’s inequality (for bounded zero-mean i.i.d. variables).
Let 
$$
\sigma^2 := \operatorname{Var}(Z_{ij}) \le \mathbb{E}[Z_{ij}^2] \le K^2 \mathbb{E}[w_i^2] \le K^2 B^2
.
$$
Then for any $t > 0$:
$$
\mathbb{P}\left( \left| \sum_{i=1}^n Z_{ij} \right| \ge nt \right) \le 2 \exp\left( - \frac{n t^2 / 2}{ \sigma^2 + M t / 3} \right).
$$
Thus, for
$
t = C \sqrt{\frac{\log(p/\delta)}{n}} \, \,\text{with } C > 0,
$
we have
$$
\mathbb{P} \left( \bigg|  \left[\nabla L(\beta^*)\right]_j \bigg| 
\ge 
(\nu + 1) C \sqrt{\frac{\log(p/\delta)}{n}} \right) 
\le 
\frac{\delta}{p}
,
$$
where,
$$
 \left[\nabla L(\beta^*)\right]_j = -\frac{\nu + 1}{n} \sum_{i=1}^n Z_{ij}.
$$

\noindent Step 4: Union bound over $j = 1, \dots, p$.
\\
Applying the union bound over all $p$ coordinates:
$$
\mathbb{P}\left( \|\nabla L(\beta^*)\|_\infty \ge (\nu + 1) C \sqrt{\frac{\log(p/\delta)}{n}} \right) \le \delta.
$$
Hence, with probability at least $1 - \delta$,
$$
\|\nabla L(\beta^*)\|_\infty \le c_1 \sqrt{\frac{\log(p/\delta)}{n}},
$$
where $c_1 >0$ depends only on $\nu, K$. The proof is completed.
    
\end{proof}

\begin{lemma}[Appendix B.3 in \cite{loh2021scale}]
\label{lm_GRSC}
Assume that Assumption \ref{assum_main} holds. Then, there exist some constant $\kappa > 0$ such that
$$
L(\beta^* + \Delta) - L(\beta^*) - \langle \nabla L(\beta^*), \Delta \rangle \ge \kappa \|\Delta\|_2^2,
$$
for all $ \| \Delta\|_2 \leq r $ and  $ \| \Delta_{S^c}\|_1 \leq 3 \| \Delta_{S}\|_1 $.
\end{lemma}

\end{document}